# Temporal Modes in Quantum Optics: Then and Now

Michael G. Raymer[1] and Ian A. Walmsley[2]
(November 2019)

[1] Department of Physics, and Oregon Center for Optical, Molecular & Quantum Science, University of Oregon, Eugene, OR 97403, USA

[2] Office of the Provost and Department of Physics, Imperial College London, South Kensington Campus, SW7 2AZ, UK

**Abstract**
We review the concepts of temporal modes (TMs) in quantum optics, highlighting Roy Glauber's crucial and historic contributions to their development, and their growing importance in quantum information science. TMs are orthogonal sets of wave packets that can be used to represent a multimode light field. They are temporal counterparts to transverse spatial modes of light and play analogous roles – decomposing multimode light into the most natural basis for isolating statistically independent degrees of freedom. We discuss how TMs were developed to describe compactly various processes: superfluorescence, stimulated Raman scattering, spontaneous parametric down conversion, and spontaneous four-wave mixing. TMs can be manipulated, converted, demultiplexed, and detected using nonlinear optical processes such as three-wave mixing and quantum optical memories. As such, they play an increasingly important role in constructing quantum information networks.

## 1. Introduction

Quantum optics is built on the concept of photons, which may be thought of as quantum particles or, more precisely, as excitations of the electromagnetic field. Roy Glauber did more than anybody to show us how to make the connection between these seemingly disparate views of quantum light. In particular, he and his collaborators formulated the concepts of coherence in quantum optics based on the mode structure and excitation of the quantized field. As a part of this work, he introduced, with U. M. Titulaer, the idea of 'temporal modes,' which are useful in describing the properties of broad-band light such as ultrashort pulses.

This paper reviews the development of the concept of 'temporal modes' (TMs) in quantum optics. TMs are orthogonal sets of wave packets that can be used to represent a multimode light field. These modes are increasingly important in the emerging fields of quantum information science and technology, for use in synchronized networks for communication and distributed computing. For example, quantum optical memories operate efficiently only if the temporal shape of the incoming light pulse matches the 'natural mode' defined by the properties of the memory.

The paper begins with the historical development of coherence in quantum optics, emphasizing Glauber's crucial contributions and the fact that the concept of spatial-temporal modes played a role all along. The properties of superfluorescence and stimulated Raman scattering were shown to be understood best using such modes as a basis. The idea of temporal modes arises when the spatial-temporal modes can be separated into spatial and temporal aspects, such as in a collimated beam or in a waveguide. The optical interactions of interest here are best understood by considering the coupling between sets of oscillators, either all-optical or also involving material systems. We classify these optical interactions as either beam-splitter-like processes or gain-like processes. Examples of beam-splitter-like processes include as optical memories, and frequency conversion by sum-frequency generation, by difference-frequency generation, or by four-wave mixing. The Schmidt-mode decomposition is used extensively to find the joint (bipartite) sets of independent mode pairs in each of these cases. We discuss the ability of pulsed frequency conversion to be temporal-mode selective, a capability that holds promise for applications in quantum information science and technology, as it provides the ability to perform quantum measurements in a temporal-mode basis. We next discuss gain-like processes, using spontaneous parametric down conversion (SPDC) and two-mode squeezing as prime examples. We conclude by emphasizing that the concept of temporal modes of the electromagnetic field, introduced by Roy Glauber and his colleagues, has proven to have significant utility both in fundamental quantum optics and in future quantum photonic applications.

## 2. Temporal Modes in Quantum Optics

Physicists began struggling seriously with the concept of quantum coherence of electromagnetic (EM) modes following the Hanbury-Brown and Twiss demonstration of stellar interferometry and the corresponding tabletop experiments devised to clarify what was 'going on.' (Hanbury Brown and Twiss 1956a) The concepts of spatial and temporal coherence had to be explicated in detail in order to understand the statistics of signals from the photoemissive light detectors being used in these experiments. (Hanbury Brown and Twiss 1956b) These efforts led, directly or indirectly, to Glauber's masterful development of quantum optical coherence theory (Glauber 1963) in the context of significant work in this area by others, such as Ugo Fano (1961), George Sudarshan (1963), and Leonard Mandel and Emil Wolf (1963).

For the purpose of studying freely propagating radiation, EM modes are complete sets of solutions of the Maxwell equations. Modes are defined conventionally in terms of four degrees of freedom: one polarization and three spatial. The most familiar mode decomposition employs monochromatic solutions of the Maxwell equations, each with a frequency $\omega$, which of course can be degenerate. The earlier heuristic concept of field quantization, by Planck and Einstein, considered these modes as harmonic oscillators whose energies were quantized in discrete amounts $\hbar\omega$. As the idea of the 'photon' became more prevalent, the usual practice was to associate a given photon with a precise amount of energy $\hbar\omega$. This kind of photon is a mathematical entity, very useful for forming a set of complete basis states to solve theoretical problems.

In the real world, the concept of a monochromatic photon (excitation of the EM field) is never exactly valid. For example, when an atom emits a photon spontaneously, the finite duration of the exponentially shaped wave packet endows a spectral width to the photon—roughly the inverse of the packet duration. Thus arises the idea of a photon 'occupying' a classically defined spatial-temporal wave packet.

The precise idea of 'spatial-temporal modes,' as they are now called, was introduced formally by Titulaer and Glauber (1966), who showed that, for the purpose of field quantization, modes need not be monochromatic. They wrote, "We note that no restrictions whatever are placed upon the spectral properties of the state $|1phot\rangle$; any pure one-photon wave packet will do, whatever its frequency distribution may be." In a beam-like geometry, the four degrees of freedom may be listed as one polarization, two spatial and one spectral; or as one polarization, two spatial and one temporal (with a Fourier transform connecting these two descriptions).

In the nineteen seventies the concept of spatial-temporal modes (STMs) reemerged in analyzing experiments on superfluorescence from collections of many two-level atoms prepared in their excited states. (Skribanowitz, *et al* 1973) See Fig. 1(a). Again, as in the Hanbury-Brown and Twiss experiments, the issue involved the *statistics* of quantum optical fields, requiring understanding the compromise between optical coherence and quantum fluctuations. (Gross, *et al* 1976; Gibbs, *et al* 1977) In a series of papers, Glauber and colleagues, notably Fritz Haake, used the so-called operator Maxwell-Bloch equations to treat the coupling of the quantized optical field to a spatially distributed set of atoms. (Glauber and Haake 1990; Haake, *et al* 1979 )

While others pursued equivalent approaches at the time (Polder, *et al* 2079), Glauber and colleagues clearly emphasized the "dominance of long-wavelength fluctuations" in the superfluorescence process, which effectively selects from the infinite variety of quantum vacuum or zero-point fluctuations a subset that couples most strongly to the evolving atom-field state. Treating superfluorescence from a long, thin medium so transverse spatial effects are unimportant, Haake et al (1979) used a semiclassical stochastic model to simulate individual superfluorescence pulses initiated by spontaneous emission. Figure 1(b) shows their numerical simulations of several optical pulses generated when a collection of atoms, all initially in their excited states, freely evolves by spontaneous and stimulated emission to create optical pulses much shorter than the spontaneous lifetime.

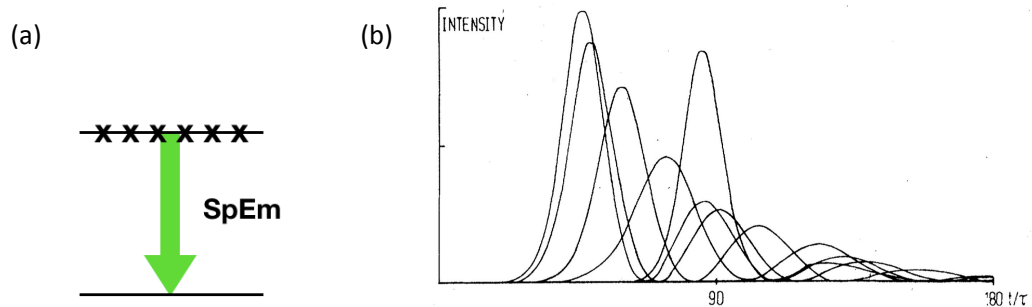

Fig. 1. a) Spontaneous emission from a collection of excited-state atoms evolves into superfluorescence, showing fluctuations (b) in the generated temporal pulse shapes. Fig. (b) reprinted, with permission, from (Haake, *et al* 1979). Copyright (1979) by the American Physical Society.

The amplitude fluctuations in superfluorescence are an example of a macroscopic quantum-noise effect. The pulses emitted by such processes are random – each realization produces a different pulse. Nevertheless, the set of pulses shown in Fig. 1(b) can each be represented as a superposition of a fairly small number of underlying orthogonal STMs acting as a mode basis, each weighted by a random amplitude. The dominance of long-wavelength fluctuations in superfluorescence gives rise to the idea of spatial-temporal modes as a natural basis for describing such processes, in that a smaller set of modes is needed when using such STMs than if using monochromatic modes.

In fact, the TM description affords the most compact means of representing a (classical) stochastic process of this kind. The Karhunen-Loève Theorem posits that any complex random process (analytic signal), e.g. the electric field $E(t)$, with a positive definite correlation function averaged over an ensemble, $C(t,t') = \langle E^*(t)E(t')\rangle$, can be expressed in terms of its eigenfunctions, $\phi_i(t)$, with weights that are uncorrelated random variables. (Saleh 1978) Thus, if $E(t) = \sum_j^\infty a_j \phi_j(t)$, with the temporal mode functions orthogonal in the time domain and the mode amplitudes uncorrelated, i.e., $\langle a_j^* a_i \rangle = \langle a_j^* a_j \rangle \delta_{ij}$, then the mode functions are defined by:

$$\int C(t,t')\phi_i^*(t')dt' = \langle a_j^* a_j \rangle \phi_i^*(t) \qquad (1)$$

where $\langle a_j^* a_j \rangle$ is the mean energy in a given temporal mode. This set of functions provides the most efficient means to represent such a time non-stationary random process, since it ensures that the mean square error between the actual field and a finite sum over the modes is minimized.

An even clearer example of STMs arising naturally in the evolution of a quantum optical field coupled to an amplifying medium is stimulated Raman scattering (SRS). Following earlier work by von Foerster and Glauber, Jan Mostowski and colleagues adapted the Glauber-Haake quantum formalism to the problem of SRS, beginning a new era in the study of fluctuations and coherence in Raman scattering, helping push the concept of temporal modes to prominence. (Mostowski and Raymer 1981; Raymer and Mostowski 1981) For review see (Raymer and Walmsley 1990).

In SRS, all atoms (or molecules) begin in their ground states and a strong laser pulse couples this ground state, via inelastic light scattering, to a state lying energetically just above the ground state, for as long as the laser pulse is present. See Fig. 2. In effect, SRS is analogous to superfluorescence, with the novelty that the laser pulse acts as an ON-OFF switch for the radiative emission. Thus, a short enough laser pulse leads to the emission of a 'Stokes-Raman' pulse in a single, fixed STM (rather than a sum of STMs as in the superfluorescence case). (Raymer, *et al* 1982) If the scattering medium has a long, thin (pencil-like) shape, thus negating transverse propagation effects, the quantum field operator at the medium's exit face at *z = L* can be approximated using a single naturally favored STM, denoted $\mathbf{v}_j(L,t)$, and photon creation and annihilation operators:

$$\hat{\mathbf{E}}(L,t) \propto \hat{A}_j \mathbf{v}_j(L,t) + \hat{A}_j^\dagger \mathbf{v}_j^*(L,t) \tag{2}$$

where the operators satisfy the standard boson commutation $[\hat{A}_j, \hat{A}_j^\dagger] = 1$. Because the STMs are functions of only time in this case, they were called 'temporal coherence modes' (Walmsley and Raymer 1986), or simply 'temporal modes' (TM). (Raymer, *et al* 1989)

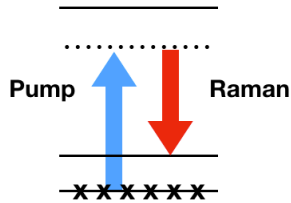

Fig. 2. Raman process, in which atoms or molecules in their ground states scatter light inelastically, leaving them in an excited state.

Because both the field and the atoms begin in their ground states, the statistics of the quantum vacuum or zero-point fluctuations determine the statistics of the generated Raman pulse. Consistent with the known properties of the quantum vacuum, the generated Raman pulse has Gaussian statistics and so can be modeled semi-classically as a single TM weighted by a complex Gaussian random variable $A$:

$$\mathbf{E}(L,t) \propto A \mathbf{v}_j(L,t) + A^* \mathbf{v}_j^*(L,t) \tag{3}$$

Thereby, each short Raman pulse is predicted to have a constant optical phase throughout its duration, although that phase (the argument of the complex number *A*) is uniformly random and unpredictable shot to shot. And the pulse energy, in this approximation, is predicted to be:

$$W \propto |A|^2 \qquad (4)$$

proportional to the time integral of $|\mathbf{v}_j(L,t)|^2$. Given that $A$ has a Gaussian probability density, it is immediately clear that the probability density of the pulse energy is exponential (Raymer, *et al* 1982)

$$P(W) = \langle \bar{W} \rangle \exp(W/\bar{W}) \qquad (5)$$

Such a probability density was observed in SRS from a hydrogen molecular gas, as shown in Fig. 3(a), illustrating a macroscopic quantum effect, in that the mean number of photons in these pulses was $10^{10}$ while the fluctuations in the number are full-scale. (Walmsley and Raymer 1983) In addition, the uniform phase distribution was verified by measuring the relative phase of interference patterns created by combining on a screen Raman pulses from two independent hydrogen-gas sources, as shown in Fig. 3(b). (Kuo, *et al* 2018)

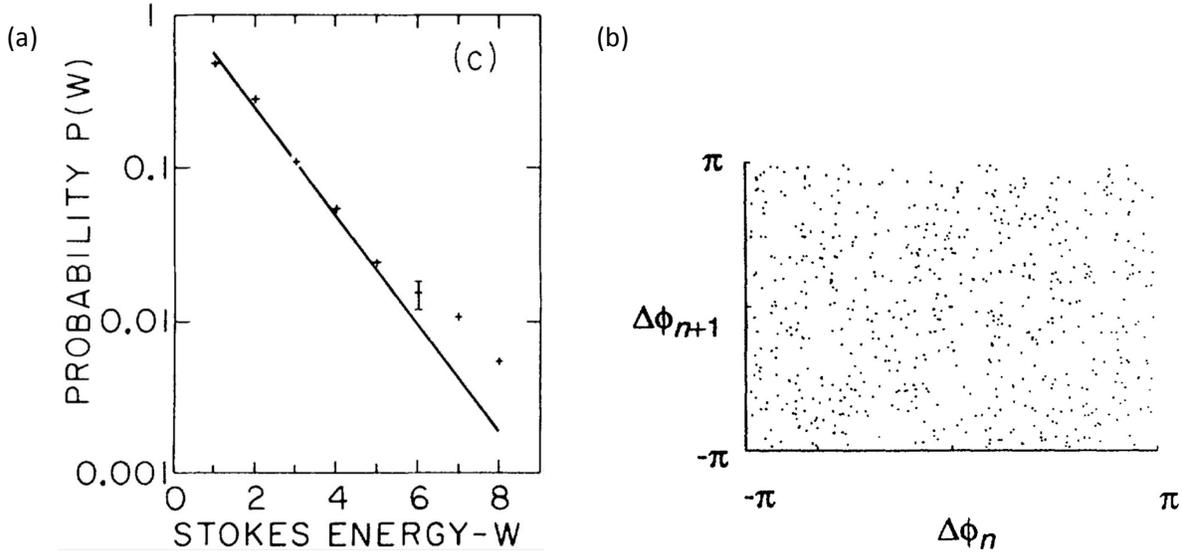

Fig. 3. a) Raman scattering pulse energy probability density, b) Relative phases between successive Raman pulses. Figs. (a, b) reprinted, with permission, from ] (Walmsley and Raymer 1983; Kuo, *et al* 2018). Copyright (1983, 1991) by the American Physical Society.

When the laser pulses driving the Raman scattering are longer than the coherence time of the medium, a sum of TMs, each with a Gaussian random amplitude, is excited in the SRS pulse, leading to shot-to-shot temporal-shape fluctuations on a fast time scale, as shown in Fig. 4(a). (Raymer, *et al* 1989) Analogous quantum fluctuations were also seen in the shot-to-shot spectrum of the Raman pulse. (MacPherson, *et al* 1988; Walmsley 1992)

The TMs relevant to this process are determined by finding the 'modes' of the Raman process, several of which are shown in Fig. 4(b). (Raymer, *et al* 1985; Walmsley and Raymer 1986; Raymer, *et al* 1989) The method for determining them is based on finding the eigenfunctions of the two-time (or two-frequency) field correlation function, as is commonly done in the Karhunen-Loève and Mercer expansions in classical coherence theory. (Saleh 1978, Wolf 1982, Wolf and Agarwal 1984) The application to composite systems, such as the Stokes pulse and the material excitation in Raman scattering, uses a generalization of these approaches - the Schmidt expansion – in which the relevant modes – the Schmidt modes – are eigenfunctions of the various self- and cross-correlation functions. We therefore refer to all such modes as 'Schmidt modes'.

In consequence of the temporal-spectral multimode nature of the Raman pulse, the total pulse energy is effectively a sum of exponentially distributed random variables, leading to a non-exponential probability density. (Raymer, *et al* 1985, Walmsley and Raymer 1986)

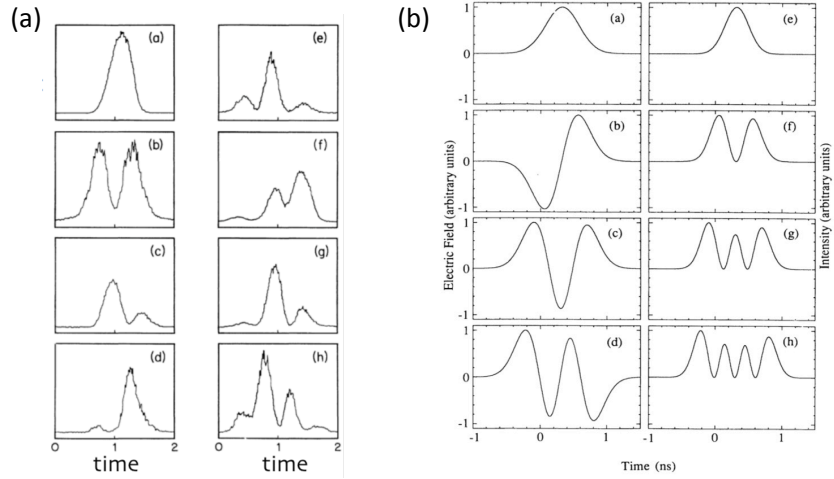

Fig. 4. a) Temporal-shape fluctuations of Stokes pulses, b) Non-exponential pulse-energy probability densities, b) Temporal modes for Stokes process. Figs. (a, b) reprinted, with permission, from (Raymer, *et al* 1989) Copyright (1989) by the American Physical Society.

There are many more recent examples of temporal modes in quantum optics—
from spontaneous parametric down conversion to quantum memories—and these will be described in the following. First, however, we explain the simplest mathematics of EM field quantization in terms of TMs.

## 3. Temporal Modes Theory—Discretizing the Continuum

The concept of temporal modes can be understood most simply in cases where light is traveling in a beam-like geometry, or in a waveguide such as an optical fiber. (Blow, *et al* 1990) Then diffraction can be either nonexistent or at least negligible, and we can focus on the longitudinal propagation of the EM field in a coordinate labeled *z*. For a single polarization, the field operator has a positive-frequency part:

$$\hat{E}^{(+)}(z,t) = i\int d\omega \sqrt{\hbar\omega/2\varepsilon_0}\, \hat{a}(\omega)e^{-i\omega(t-z/c)} \qquad (6)$$

where *c* is the speed of light. The monochromatic-mode annihilation and creation operators satisfy the boson commutator $[\hat{a}(\omega), \hat{a}^\dagger(\omega')] = \delta(\omega-\omega')$. A linear superposition of these creation operators defines a discrete photon creation operator $\hat{A}_j^\dagger$

$$\hat{A}_j^\dagger = \int d\omega\, f_j(\omega)\hat{a}^\dagger(\omega) \qquad (7)$$

If a set of orthogonal weight functions $f_j(\omega)$ is used to define a set of such creation operators, then the new operators obey the standard boson commutation relation $[\hat{A}_j, \hat{A}_k^\dagger] = \delta_{jk}$. Here orthogonality is defined by $\int d\omega\, f_j^*(\omega) f_k(\omega) = \delta_{jk}$, and completeness by $\sum_j f_j^*(\omega) f_j(\omega') = \delta(\omega - \omega')$. Then each $\hat{A}_j^\dagger$ creates one excitation in a particular TM denoted as $v_j(z,t)$, defined by (Titulaer and Glauber 1966)

$$v_j(z,t) = i\int d\omega \sqrt{\hbar\omega/2\varepsilon_0}\, f_j(\omega) \exp[-i\omega(t-z/c)] \qquad (8)$$

These TMs serve as a mode basis for the field operator via (Titulaer and Glauber 1966)

$$\hat{E}^{(+)}(\mathbf{r},t) = \sum_j \hat{A}_j v_j(z,t) , \qquad (9)$$

where we used the inverse relation $\hat{a}(\omega) = \sum_j f_j(\omega) \hat{A}_j$, which follows from the completeness condition.

The TMs have the curious property of being non-orthogonal under the usual spatial overlap integral:

$$\int v_j^*(z,t) v_k(z,t)\, dz \neq 0 \quad \text{for } j \neq k \qquad (10)$$

Nonorthogonality is a consequence of the factor $\sqrt{\hbar\omega}$ in Eq.(8), making it a nonunitary transformation. Fortunately, when the modes of interest are composed only of spectral components centered reasonably close to a central carrier frequency, the modes are very nearly orthogonal. This holds for pulses with durations even as brief as a few optical cycles. Within this approximation the functions $v_j(z,t)$ form an orthogonal basis for representing the spatial-temporal structure of the EM field within a given narrow spectral band. In this case, the temporal modes in the time domain are proportional to the Fourier transforms of the $f_j(\omega)$, denoted as $\tilde{f}_j(t)$.

In noninteracting free-space propagation, there is no preferred basis choice. An example of such a set of orthogonal TMs is the set of Hermite-Gaussian functions, and their spectral equivalents, a few of which are illustrated in Fig. 5.

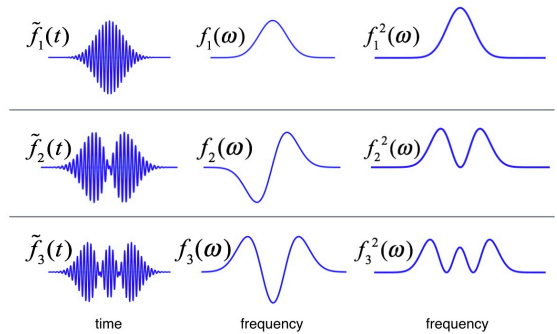

Fig. 5 Examples of a basis set of temporal modes. left: Temporal field, middle: Spectral amplitudes, right: Spectral intensities.

Temporal modes are a basis for the classical Maxwell equations, and form their own Hilbert space with properties that have analogs to the quantum state space of excitations.

Although it is tempting to call $v_j(z,t)$ the 'state' of the photon, such language can be misleading if used without care. (Smith and Raymer 2007) In particular, it does not easily apply to quantum states of the field beyond individual photons. (Van Enk 2007, Lamb 1995) In this sense, a mode, including a temporal mode, is like a 'container' into which arbitrary states can be placed.

As stated earlier, a one-photon state of a given TM is

$$|1\rangle_j = \hat{A}_j^\dagger |vac\rangle \tag{11}$$

where $|vac\rangle$ is the vacuum state of *all* EM modes.

A given TM may also be excited into a Glauber coherent state:

$$|\alpha\rangle_j = \exp\left(\alpha \hat{A}_j^\dagger - \alpha^* \hat{A}_j\right)|vac\rangle \tag{12}$$

or into a squeezed-vacuum state:

$$|\xi\rangle_j = \exp\left(\xi \hat{A}_j^{\dagger 2} - \xi^* \hat{A}_j^2\right)|vac\rangle \tag{13}$$

Similar expressions may be written for multi-mode excitations, leading, for example, to quantum entanglement between excitations in different field modes.

The key point of this approach is that the continuously infinite collection of monochromatic modes has been replaced by a countably infinite set of discrete modes, only a small number of which may be needed to describe a particular situation. Further details of the STM construction, including the three-dimensional treatment, are expounded in (Smith and Raymer 2007, Titulaer and Glauber 1966).

## 4. Bipartite Oscillator Interactions

When two quantum-oscillator systems are coupled, their joint state may become nonseparable and may therefore be used as an entangled-state resource for quantum information. Both oscillator systems may be optical field modes, or one may be an optical field and the other a large collection of weakly excited atoms or molecules whose collective excitation can be approximated as an effectively harmonic system with equal energy-level spacings. Examples of the latter include atomic 'spin waves' in low-density vapors and vibrational excitation waves (optical or acoustic phonons) in materials—molecular gases, liquids or solids. The spin-wave case is important in many realizations of optical quantum memories implemented through 'slow light' or electromagnetically induced transparency. (Liu, *et al* 2001, Boller, *et al* 1991, Fleischhauer and Lukin 2000) In the case of molecular gases or solids, such delocalized vibrations can be created by collective Raman scattering, for example. (Simon, *et al* 2010), Lee, *et al* 2011)

It is useful to think in terms of scattering processes, illustrated in Fig. 6. The 'systems' may be optical or material. The 'pumps' are typically laser light pulses that create coupling and provide energy to or take energy from the system. Sets of initial (input) mode operators $(\hat{a}_n, \hat{b}_n)$ evolve to final (output) mode operators $(\hat{c}_n, \hat{d}_n)$. In Fig. 6(a) both systems are optical fields, which interact through a 'passive' nonlinear-optical medium, whose state does not change, driven by one or more pumps. In Fig. 6(b) one system is optical, while the other is an 'active' medium, whose state may change during the interaction.

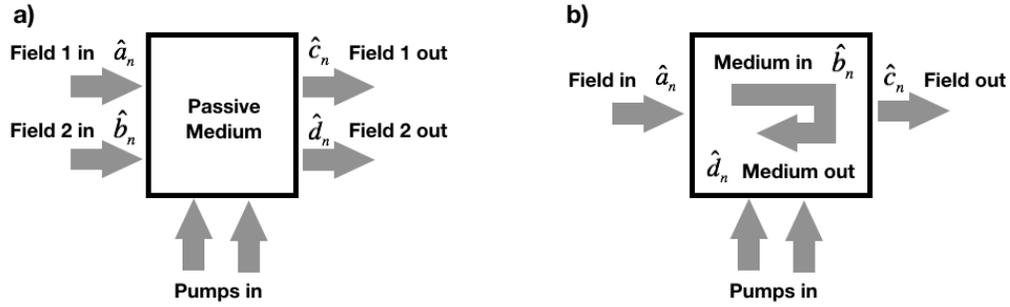

Fig. 6 Nonlinear optical process, with two oscillator systems evolving from initial (input) to final (output) conditions in the presence of laser 'pump' pulses. a) represents optical parametric processes where the medium ends up in the same state in which it began; b) represents a light-matter coupling that leaves the medium in a state different than where it started.

There are two classes of bipartite oscillator interactions: *beam-splitter-like process*es and *gain-like processes*. A beam-splitter-like process is inherently background free. That is, if both of the oscillators are initially in their ground (vacuum) states, they remain so. And if one or both oscillators are excited initially, then these excitations can be swapped between them. The simplest example is represented by the two-mode transformation relating input and output operators (choosing transmission and reflection amplitudes $\tau, \rho$ to be real):

$$\hat{a} = \tau\hat{c} - \rho\hat{d}, \quad \hat{b} = \rho\hat{c} + \tau\hat{d} \qquad (14)$$

where $\tau^2 + \rho^2 = 1$. Note that there is no mixing of creation and annihilation operators, so there is no squeezing or spontaneous generation of photons.

Figure 7 illustrates examples of beam-splitter-like processes. Figures 7(a) through 7(d) correspond to Fig. 6(a), while Fig. 7(e) is an example of the process in Fig. 6(b), where the medium's state changes. These processes are discussed in more detail below.

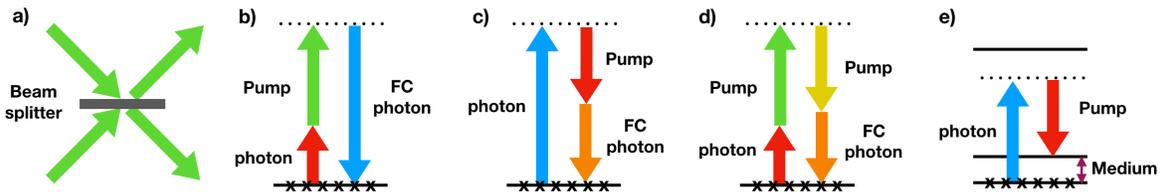

Fig. 7 (a) Ordinary beam splitter, and four beam-splitter-like processes: frequency conversion by: (b) sum-frequency generation; (c) frequency conversion by difference-frequency generation; (d) frequency conversion by four-wave mixing; (e) and optical memory. Solid lines are 'real' states and dotted lines are 'virtual' states. The conjugates of these processes are also beam-splitter-like processes.

In the Heisenberg picture, the operators evolve but the state does not. For example, if a single-photon state $\hat{a}^\dagger |vac\rangle_{ab} = |1\rangle_a |0\rangle_b$ representing a single excitation in system *a*, and no excitation in system *b*, is incident on an ordinary optical beam splitter, so that the two systems are two optical modes, as in Fig. 6(a), the two systems are coupled. The state expressed in terms of the output-mode operators is

$$|1\rangle_a|0\rangle_b = (\tau \hat{c}^\dagger - \rho \hat{d}^\dagger)|vac\rangle_{cd} = \tau|1\rangle_c|0\rangle_d - \rho|0\rangle_c|1\rangle_d \qquad (15)$$

where we used the fact that the vacuum remains invariant under the beam-splitter transformation. This state shows that the single-photon excitation is shared by the two outgoing modes, whose fields are now in an entangled state. (Van Enk 2007) Equation (15) can be viewed as merely a 'classical' mode transformation, which in quantum theory amounts to a (global) change of basis. If, instead, one of the systems is a material one, as in Fig. 6(b), then the states $|0\rangle_b, |1\rangle_b$ (as well as $|0\rangle_d, |1\rangle_d$) indicate zero or one material excitation in that system.

In contrast, in a *gain-like process*, there is the possibility of stimulated emission or scattering. Therefore, spontaneous emission is also present, meaning that even if there is no initial excitation in either of the oscillator systems, they can become excited. (Caves 1982) A familiar example is a laser amplifier. It will emit light into an output mode even if there is no input – this process is termed amplified spontaneous emission (ASE). ASE is usually considered as background noise, which for single-excitation input states can be substantial.

Gain-like processes are described by the well-known two-mode squeezing transformation, studied by Mollow and Glauber (1976a, 1976b), among others. This leads to the input-output relation:

$$\hat{a} = \mu \hat{c} + \nu \hat{d}^\dagger, \quad \hat{b} = \mu \hat{d} + \nu \hat{c}^\dagger \qquad (16)$$

where $\mu^2 - \nu^2 = 1$, which mixes creation and annihilation operators. This transformation arises from a Hamiltonian of the form $\hat{H} = -i\varepsilon \hat{c}\hat{d} + i\varepsilon \hat{c}^\dagger \hat{d}^\dagger$, where $\varepsilon$ depends on the pump powers and medium parameters. For example, the vacuum evolves, to leading order, to

$$|0\rangle_c|0\rangle_d \to \exp(-i\hat{H})|0\rangle_c|0\rangle_d \approx \sqrt{1-\varepsilon^2}|0\rangle_c|0\rangle_d + \varepsilon|1\rangle_c|1\rangle_d + ... \qquad (17)$$

where we have taken $\hbar$ to be unity for simplicity. This expression illustrates the fact that such processes generate excitations spontaneously. In this case, the number of excitations in the two output modes are equal, and the superposition of these correlated photon-number components means that the state describing the joint system is non-separable.

For systems with more than two input and output modes, it is important to identify which modes interact via these interactions, and how the resulting excitations are distributed throughout the multimode oscillator systems. Most importantly, we want a concise way to specify which modes become entangled with which other modes. Is the entanglement bipartite or multipartite? A temporal-mode analysis answers these questions, in several different contexts, discussed next.

## 5. Quantum Frequency Conversion

A prime example of a beam-splitter-like process is quantum frequency conversion (FC), which occurs when the state of a given spectral region (or 'band'), for example red, is transferred all or in part to a band centered at a different frequency, say blue. (Kumar 1990) Because quantum evolution is unitary, the states in these two bands are actually swapped. A simple way to visualize this process is to consider a rapidly moving mirror or beam splitter, which blue-shifts light reflected from the 'front' of the beam splitter and red-shifts light reflected from the 'back.' (Raymer, *et al* 2010) Because the frequency shifts are equal and opposite, if the reflectively is 100% the colors defining the modes are 'swapped' while their quantum states are preserved. For example, if the input is the Fock state $|3\rangle_{red}|2\rangle_{blue}$, it can be transformed into $|2\rangle_{red}|3\rangle_{blue}$. In the case of partial FC, the same input state can be transformed into the entangled state $|3\rangle_{red}|2\rangle_{blue} \pm |2\rangle_{red}|3\rangle_{blue}$. In both cases, the number of photons in the two signal fields is preserved while the energy is not (the additional energy is provided by the time-non-stationary linear coupling – the moving beamsplitter in the above example).

Quantum frequency conversion across large frequency shifts was first implemented using nonlinear-optical sum-frequency generation, as illustrated in Fig. 7(b). (Huang and Kumar 1992) A pump laser pulse, shown as green, loses a photon, whose energy is added to that of the red photon, creating a blue photon. A similar process—difference-frequency generation—is shown in Fig. 7(c), where the signal photon gives up some energy to the pump pulse, decreasing its frequency. (Ding and Ou 2010) More flexible FC across large or small frequency shifts can be achieved using nonlinear-optical four-wave mixing in third-order nonlinear materials such as silica glass, as illustrated in Fig. 7(d). (McKinstrie, *et al* 2005) Two pump fields interact with the signal to increase or decrease the input photon's frequency by the difference of the two pumps' frequencies, as first reported using photonic-crystal optical fiber in McGuinness, *et al* (2010a, 2010b). For quantum FC over very small frequency ranges ($\leq$ 100 GHz), direct phase modulation by an electrooptic crystal can be used. (Wright, *et al* 2017)

The details of the nonlinear optical process depend on the temporal shape of the laser pulse or pulses used to drive the process as well as the phase-matching conditions defined by the medium's dispersion. In general, FC between bands is not complete but partial, and the process is described using a generalization of Eq.(14); for review see (Raymer, *et al* 2010, Christ, *et al* 2013)

$$\hat{a}(\omega) = \int G^{ac}(\omega,\omega')\hat{c}(\omega')d\omega' + \int G^{ad}(\omega,\omega')\hat{d}(\omega')d\omega'$$
$$\hat{b}(\omega) = \int G^{bc}(\omega,\omega')\hat{c}(\omega')d\omega' + \int G^{bd}(\omega,\omega')\hat{d}(\omega')d\omega' \quad (18)$$

where $\hat{a}(\omega)$ and $\hat{c}(\omega)$ refer to one frequency band (say 'red') and $\hat{b}(\omega)$ and $\hat{d}(\omega)$ refer to the other band ('blue'). The Green functions $G^{ij}$ are elements of a unitary transformation, and thus obey, for example,

$$\int d\omega' \left\{ G^{ac}(\omega_1,\omega')G^{ac*}(\omega_2,\omega') + G^{ad}(\omega_1,\omega')G^{ad*}(\omega_2,\omega') \right\} = \delta(\omega_1 - \omega_2) \quad (19)$$

The method of temporal modes leads to the clearest and most compact description of FC. Performing a Schmidt (Mercer) decomposition of each of the Green functions that describe the FC process, a derivation similar to the 'Bloch-Messiah theorem' (Braunstein 2005) leads to pleasing relationships between their forms (Raymer, *et al* 2010)

$$G^{ac}(\omega,\omega') = \sum_n \tau_n V_n(\omega) v_n^*(\omega')$$
$$G^{bd}(\omega,\omega') = -\sum_n \rho_n V_n(\omega) w_n^*(\omega')$$
$$G^{ad}(\omega,\omega') = \sum_n \rho_n W_n(\omega) v_n^*(\omega') \quad (20)$$
$$G^{bc}(\omega,\omega') = \sum_n \tau_n W_n(\omega) w_n^*(\omega')$$

Rather than needing eight separate sets of Schmidt-mode functions, just four such sets are sufficient to describe the dynamics, as a result of the constraints imposed by unitarity. The coefficients act like beam-splitter amplitudes, with $\tau_n$ playing the role of a transmission coefficient (not changing the color) and $\rho_n$ the corresponding reflection coefficient (changing the color), for each TM. For example, $\rho_n W_n(\omega) v_n^*(\omega')$ refers to the process in which a photon in the temporal mode $v_n(\omega')$ within the initial frequency band is converted into a photon in the mode $V_n(\omega)$ in the final band, with probability $\rho_n^2$. And $\tau_n V_n(\omega) v_n^*(\omega')$ refers to the process in which a photon in the mode $v_n(\omega')$ is converted into a photon in the temporal mode $V_n(\omega)$ in the same frequency band, with probability $\tau_n^2$.

Using this Bloch-Messiah-Schmidt decomposition, as we will call it, the input and output operators in Eq.(18) are expressed as:

$$in: \quad \hat{a}^\dagger(\omega) = \sum_n \hat{A}_n^\dagger V_n(\omega), \quad \hat{b}(\omega) = \sum_n \hat{B}_n^\dagger W_n(\omega)$$
$$out: \quad \hat{c}^\dagger(\omega) = \sum_n \hat{C}_n^\dagger v_n(\omega), \quad d(\omega) = \sum_n \hat{D}_n^\dagger w_n(\omega) \tag{21}$$

where $\hat{A}_n^\dagger, \hat{B}_n^\dagger, \hat{C}_n^\dagger, \hat{D}_n^\dagger$ are discrete bosonic operators for the TMs defined by the FC process, obeying $[\hat{A}_n, \hat{A}_m^\dagger] = \delta_{nm}$, etc. The remarkable feature of this decomposition is that every FC interaction occurs only between two input TMs and two output TMs, represented by the beam-splitter relations, analogous to Eq.(14):

$$\hat{A}_n = \tau_n \hat{C}_n - \rho_n \hat{D}_n, \quad \hat{B}_n = \rho_n \hat{C}_n + \tau_n \hat{D}_n \tag{22}$$

Thereby, a FC process couples TMs before and after conversion in a strictly pair-wise manner, greatly simplifying the original continuous-frequency description. That is, if a photon comes into the FC process in a particular TM of one color, it will be converted to a particular TM of the other color. The TMs do not mix in the process. By careful design of the physical configuration (for example, the phase and group velocities of the waves in the nonlinear medium) it is possible to arrange that any input TM is mapped perfectly to the corresponding output TM at the upconverted (or downconverted) frequency.

Because FC is beam-splitter-like, it is fundamentally background free, and preserves the quantum state of the light. (Tanzilli, *et al* 2005, McGuinness, *et al* 2010a, Rakher, *et al* 2010) It therefore provides a significant functionality for quantum information science. (Raymer and Srinivasan 2012, Brecht, *et al* 2015)

**6. Quantum Optical Memory**

Another important beam-splitter-like process is quantum optical memory: a means of storing coherently the state of a traveling optical mode in a stationary collection of atoms or molecules. The state could be a Fock state, a coherent state, a squeezed state, or a superposition of any of these, provided the maximum number of photons is much less that the number of atoms or molecules, so that the harmonic oscillator approximation holds. Later the same state can be 'read out' into a newly generated optical pulse. The read-in and read-out processes are controlled by the pump pulse(s). (Lukin 2003, Cirac, *et al* 2004) For example, a memory based on far-off-resonance Raman scattering is shown in Fig. 7(e); a single-photon wave packet (blue) is absorbed while a photon is added to the pump pulse, leaving the medium in a state with one excitation distributed coherently throughout the medium. (Nunn, *et al* 2007) The pump pulse, being strong and coherent, is negligibly affected. Here, too, the concept of TMs plays a powerful role, as we now explain.

The optical memory that is most closely analogous to quantum frequency conversion, discussed above, is Brillouin scattering. This process is similar to Raman scattering, but involves *acoustic* phonons in a crystal, glass or liquid medium, which propagate at the speed of sound. (Garmire 2018) Zhu et al pointed out that optical pulses can be transferred coherently into the material vibrations, creating an optical memory. (Zhu, *et al* 2007) The interaction Hamiltonian for Brillouin scattering is formally the same as for sum or difference-frequency generation, so the operator transformation is the same as in Eq.(18), where here the operators $(\hat{b}_n, \hat{d}_n)$ refer to the medium, and $(\hat{a}_n, \hat{c}_n)$ refer to the optical signal field, as in Fig. 6(b). Therefore, we point out, the ensuing TM analysis in Eqs.(20)-(22) again hold.

Raman scattering, which involves *optical*, rather than acoustic, phonons, was described using TMs in (Raymer, *et al* 1982), Raymer, *et al* 1989, Wasilewski and Raymer 2006) In this case a slight change of formalism is needed because the phonon excitation in the medium does not propagate (the group velocity is zero for atomic spin waves or molecular or solid-state optical phonons). For the material excitation, the analog of the TM for an optical field is a longitudinal spatial mode of the medium. Otherwise the mathematics is similar to that of the FC process: the state of the incoming optical field TM can be 'written' into the corresponding spatial mode of the medium, thus creating a quantum memory. (Nunn, *et al* 2007)

The transformation is still beam-splitter-like, as there is no gain or squeezing. Define $\hat{c}(0,t)$ as the incident field annihilation operator at $z = 0$, and $\hat{d}(z,0)$ the medium's initial collective annihilation operator at $t = 0$, before interaction; likewise $\hat{a}(L,t)$ is the outgoing field operator and $\hat{b}(z,T)$ the medium's final operator after interaction. Then solutions of the Heisenberg-picture equations of motion can be represented by (Wasilewski and Raymer 2006, Nunn, *et al* 2007):

$$\hat{a}(L,t) = \int_0^T G^{ac}(t,t')\hat{c}(0,t')dt' + \int_0^L G^{ad}(z',t)\hat{d}(z',0)dz'$$
$$\hat{b}(z,T) = \int_0^T G^{bc}(z,t')\hat{c}(0,t')dt' + \int_0^L G^{bd}(z,z')\hat{d}(z',0)dz' \quad (23)$$

Here we work explicitly in the space-time variables, rather than the frequency variable as in the FC examples above, and the transformation is in the 'forward' rather than the inverse sense. Again, the Green functions together comprise a unitary transformation, analogous to a beam-splitter transformation (no mixing of creation and annihilation operators).

In this case the Bloch-Messiah-Schmidt decomposition yields a set of Schmidt modes that are mixed in space and time variables: $\psi_n^{(in)}(t), \psi_n^{(out)}(t)$ to describe the time dependence of input and output optical fields, and $\varphi_n^{(in)}(z), \varphi_n^{(out)}(z)$ to describe the initial and final spatial distributions of the collective material excitation. The four mixed space-time Green functions can be expressed in terms of the Schmidt modes analogously to Eq.(20). And the output/final operators $\hat{a}(L,t)$ and $\hat{b}(z,T)$ can be expressed analogously to Eq.(21), again with the beam-splitter-like relation between initial and final operators, of the form Eq.(22).

Again, a good way to think about these scenarios is that the dynamics (in the Heisenberg picture) amount only to a mode transformation, now for both fields and medium vibrations, equivalent to the same transformations that occur in the classical Maxwell equations and classical multimode oscillator theory. A unique set of field-and-medium modes can be obtained via the Bloch-Messiah-Schmidt decomposition, which set has the property that it is the one in which each mode is connected to as few others as is possible. The quantum field obeys the standard bosonic commutation relations among mode creation and annihilation operators.

## 7. Temporal-Mode Demultiplexing in Beam-Splitter-Like Processes

An important aspect of pulsed frequency conversion is that it is temporal-mode selective, as first pointed out for four-wave mixing by McGuiness *et al* (2011) It was proposed by Eckstein *et al* (2011) for the case of three-wave mixing that, given the shape(s) of the driving laser pulse(s), the set of Schmidt coefficients $\lambda_j$ can be designed such that only one coefficient $\lambda_0$ is large while the rest are much smaller. This occurs in the limiting case that the Green functions are separable, then only one pair of TMs are coupled; the rest pass through the FC process unscathed. Light in the input mode is converted to the output mode with probability , and vice versa. Figure 8 illustrates this process, which Silberhorn's group demonstrated and named a 'quantum pulse gate' or QPG. (Brecht, *et al* 2014) A high-efficiency version of the QPG was proposed and demonstrated in (Reddy and Raymer 2018). Other groups have further analyzed, demonstrated, and refined the development of the QPG, and discussed its applications in quantum information science. (Huang and Kumar 2013, Reddy, *et al* 2013, Reddy, *et al* 2015, Shahverdi, *et al* 2017) For reviews, see (Brecht, *et al* 2015, Ansari, *et al* 2018).

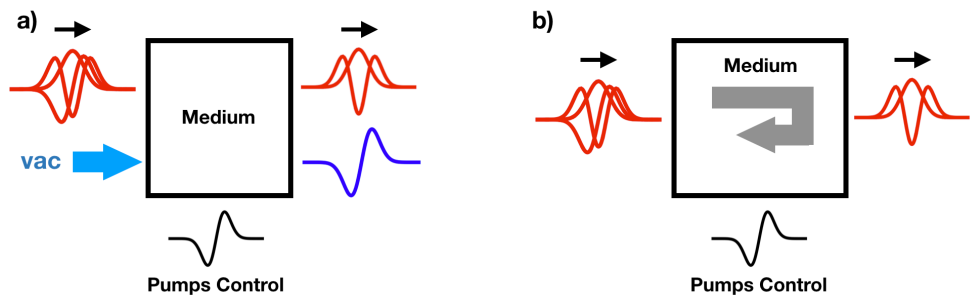

Fig. 8 The Quantum Pulse Gate (QPG). Temporal-mode selectivity and demultiplexing by: a) all-optical frequency conversion, b) medium-based optical memory.

And, given that optical memory is governed by the same underlying equations of motion as FC, it too should be TM selective. Indeed, early theories of optical memory recognized this fact to a certain extent, although usually not emphasized. (Cirac, *et al* 2004), Novikova, *et al* 2007, Gorshkov, *et al* 2007) The case of far-off-resonant Raman memory, discussed above, was discussed explicitly in the context of TM selectivity by Nunn et al. (Nunn, *et al* 2007) Namely, given the physical parameters of the Raman medium and the temporal shape of the pump pulse, only a single TM component of an arbitrary input signal pulse is trapped and stored in the memory, as illustrated in Fig. 8(b). At the same time, the information about which TM the light originally occupied is 'erased,' allowing a new TM identity to be assigned to it upon subsequent readout. In this way the memory acts not only as a selective 'pulse gate' on read-in but also as a mode transformer on readout. TM-selective storage and transformation in a quantum memory have been demonstrated. For example, a Raman memory in a warm atomic caesium vapour has shown efficient storage of a temporal 0th-order Hermite-Gaussian pulse with an inter-modal contrast of nearly 9dB. (Hird, *et al* 2018) Further, the readout was into a $1^{st}$ or $3^{rd}$-order Hermite-Gaussian pulse. Indeed a 'modal tomography' of the memory showed that it could convert between four such modes with fidelity of over 90%.

Temporal mode selectivity enables a very important functionality—TM multiplexing. In general, multiplexing of optical modes is a necessary operation in classical and quantum information systems. Familiar examples are: wavelength-division multiplexing or time-domain multiplexing in telecom systems, transverse-spatial-mode multiplexing by holographic gratings, and polarizing beam splitters for separating to orthogonal light polarizations. Pulsed FC and optical memory can both serve as multi-channel 'polarizing beam splitters' for a given set of TMs. Because FC is formally equivalent to a multichannel beam splitter, any linear-optics operations that can be done using beam splitters can also be done using FC, thereby providing a 'complete framework for quantum information science,' as pointed out in (Brecht, *et al* 2015).

### 8. Photon Pair Generation and Squeezing

Zel'dovich and Klyshko (1969) presented a quantum mechanical treatment of spontaneous parametric down conversion (SPDC), the process by which single high-frequency ('blue') photons are annihilated in a nonlinear optical crystal, spontaneously creating pairs of lower-frequency ('red') photons, as illustrated in Fig. 9. In the next decades, SPDC became a workhorse of quantum optics, because it creates quantum correlations and entanglement in the generated light, enabling studies of the Bell inequalities and the generation of 'heralded' photons, meaning that the detection of one of the pair of photons indicates the presence of the second photon in a separate beam. (Hong and Mandel 1985, Shih and Alley 1986, Ghosh and Mandel 1987, Hong, *et al* 1987, Franson 1989) The preparation of a specific TM and spatial mode by this process is also possible, with careful design of the downconverter. (Grice *et al* 2001; Garay-Palmett, et al 2007, Mosley *et al* 2008a, Cohen *et al* 2009, Brecht, *et al* 2015, Ansari, *et al* 2018)

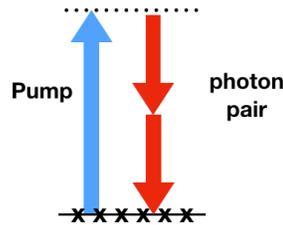

Fig. 9. Spontaneous parametric down conversion (SPDC) process.

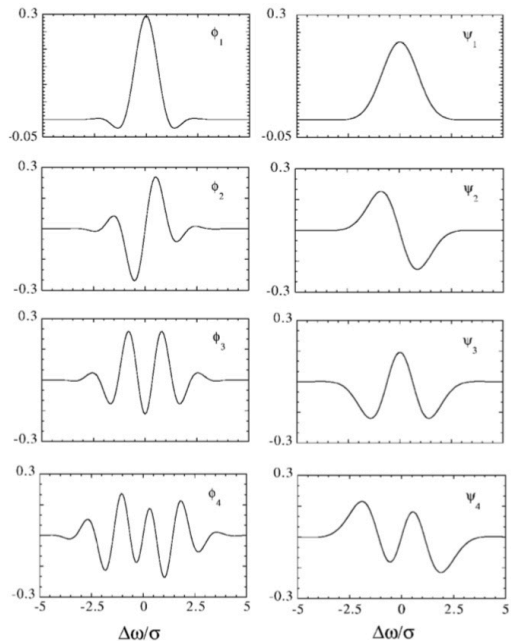

Fig. 10. The first four Schmidt-mode pairs for a typical SPDC process. From (Law, *et al* 2000)

For SPDC pumped by short laser pulses, the most effective analysis is based on a temporal mode decomposition. (Law, *et al* 2000) To model SPDC in a 'traveling-wave' geometry, we denote the (scaled) field operators by

$$\hat{E}_\alpha^{(+)}(z,t) = \int d\omega\, \hat{a}_\alpha(\omega) e^{-i\omega(t-z/c)} \tag{24}$$

for $\alpha = 1, 2$, where for simplicity we omitted any frequency-dependent prefactor $i\sqrt{\hbar\omega/2\varepsilon_0}$ in the integral of Eq.(6). In the limit of weak excitation of the down-converted field, the joint state of the two systems can be written as a generalization of Eq.(17):

$$|\Psi\rangle \cong \sqrt{1-\varepsilon^2}\,|vac\rangle + \varepsilon \iint d\omega\, d\tilde{\omega}\, \Psi(\omega,\tilde{\omega})\hat{a}_1^\dagger(\omega)\hat{a}_2^\dagger(\tilde{\omega})|vac\rangle \tag{25}$$

where $\Psi(\omega,\tilde{\omega})$ is the normalized 'two-photon amplitude,' and $\varepsilon^2\,|\Psi(\omega,\tilde{\omega})|^2$ is the joint probability for detecting one quantum in each system with frequencies $\omega, \tilde{\omega}$, respectively. Note the perfect correlation of excitation numbers—if one system contains an excitation then both do.

In general, the two-photon amplitude is not factorable in frequency ($\Psi(\omega,\tilde{\omega}) \neq f(\omega)g(\tilde{\omega})$), indicating spectral entanglement of the pair of photons, since the two photon state is not then separable. The extent and character of the nonseparability can be best understood using a 'Schmidt,' 'Mercer,' or singular-value decomposition (SVD) (Law, *et al* 2000):

$$\Psi(\omega,\tilde{\omega}) = \sum_j \sqrt{\lambda_j}\,\psi_j(\omega)\phi_j(\tilde{\omega}) \tag{26}$$

where $\psi_j(\omega), \phi_j(\tilde{\omega})$ are ortho-normal sets of 'Schmidt modes' for systems 1 and 2, respectively, with associated 'singular values' $\lambda_j$. If $\Psi(\omega,\tilde{\omega})$ is separable, meaning only one singular value is nonzero, then there is no entanglement between the systems.

In analogy to Eq.(7), we define TM creation operators by

$$\hat{A}_j^\dagger = \int d\omega\, \psi_j(\omega)\hat{a}_1^\dagger(\omega), \quad \hat{B}_j^\dagger = \int d\omega\, \phi_j(\omega)\hat{a}_2^\dagger(\omega) \tag{27}$$

and these obey the standard boson commutation relations for creation and annihilation operators, $[\hat{A}_j, \hat{A}_k^\dagger] = \delta_{jk}$, etc. In terms of the Schmidt-mode operators, the joint state is expressed as:

$$|\Psi\rangle = \sqrt{1-\varepsilon^2}\,|vac\rangle + \varepsilon \sum_j \sqrt{\lambda_j}\,\hat{A}_j^\dagger \hat{B}_j^\dagger |vac\rangle, \tag{28}$$

illustrating that the doubly-continuous frequency degree of freedom has been discretized into correlated sets of TMs.

The field operators Eq.(24) can also be expressed in terms of the Schmidt modes, defining the temporal modes in the time domain as: $u_j(t) = \int d\omega\, e^{-i\omega t}\psi_j(\omega)$ and $v_j(t) = \int d\omega\, e^{-i\omega t}\phi_j(\omega)$, which gives:

$$\hat{E}_1^{(+)}(z,t) = \sum_j \hat{A}_j^\dagger u_j(t - z/c) \qquad (29)$$

$$\hat{E}_2^{(+)}(z,t) = \sum_j \hat{B}_j^\dagger v_j(t - z/c) \qquad (30)$$

Thus, the correlation (entanglement) that is distributed in a seemingly complicated manner across the spectral continuum has been identified as existing only between pairs of discrete temporal modes. The TMs are defined uniquely by the photon-pair generation process itself. Importantly, in typical cases only a small number (say one to ten) TM pairs are required to capture the two-photon state, in contrast to the very much larger number of monochromatic frequency modes.

An example of TM pairs is shown in Fig. 10, for SPDC of a sub-ps laser pulse in a 0.8 mm-long second-order nonlinear optical crystal. (Law, *et al* 2000) In this case, the values of the four largest singular values, corresponding to the TMs shown in the figure, were 0.65, 0.19, 0.067, 0.028, respectively, with the remaining values even smaller, indicating that only five or six Schmidt modes are sufficient to capture the essence of this bipartite state. Experiments have demonstrated means to control the number of TM pairs created in SPDC. (Brecht, *et al* 2015, Ansari, *et al* 2018) Further, it is possible to filter single photons using a cavity, at the same time as they are stored by this means. This enables creation of a bright single-photon source by means of temporal multiplexing: single-photon pulses are generated in a single TM, using a repeat-until-success protocol, storing the successful events, then releasing them on demand. (Kanda, *et al* 2015)

SPDC is a gain-like process in that spontaneous and stimulated emission are present. As the pump strength is increased, more and more photon pairs can be generated, leading to 'twin beam' generation, named as such because both beams (the oscillator systems) contain equal numbers of photons. This process is also called two-mode squeezing, as the difference of the amplitudes of the beams carries a reduced level of quantum fluctuations. (Mollow 1973) Early experiments observed evidence of twin-beam generation in the reduced fluctuations (noise) of the difference of the two beam's intensities. (Heidmann, *et al* 1987), Aytür and Kumar 1990) Those experiments observed noise reduction in the radio-frequency (rf) spectrum of the intensity at frequencies typically in the 10 MHz range.

A reduction in the relative fluctuation of energy of two whole pulses (rather than an rf spectral component) was observed by Smithey *et al* (1992). To achieve this result, each entire pulse was predominantly excited in a single temporal mode (in this case defined by a short coherent 'seed' pulse).

In such cases, where the gain is high, perturbation theory no longer holds, and the analysis becomes far more complicated. The TMs that are defined in the low-gain limit by Eq.(26) are *not* the same as the TMs that govern the high-gain case. Pulse propagation and amplification leads to distortions of the TMs for high gain. (Christ, *et al* 2013, Quesada and Sipe 2015) Nevertheless, strong two-mode squeezing in predominantly single TMs has been observed in experiments with up to 20 photon pairs. (Harder, *et al* 2016)

**9. Measurements based on TMs**

An important recent innovation is the ability to perform quantum measurements in a temporal-mode basis. (Eckstein, *et al* 2011, Huang and Kumar 2013, Brecht, *et al* 2014, Brecht, *et al* 2015, Reddy and Raymer 2018) Measurements on optical fields usually consist of absorption of optical energy into a material and the subsequent measurement of electric currents, which may be, for example, from photoexcited charges or from a disturbed superconducting current. Typically, a spatially extended photodetector absorbs energy from an optical field over a particular bandwidth, with a characteristic response time. It is important to take into account how the detector responds to different modes of the optical field, especially in cases where these are not the same as the modes into which information is encoded in the quantum light field, or the natural modes of the detector. This is particularly important for quantum optical sensing and imaging, and applies also to quantum communications and even quantum computation and simulation.

Huang and Kumar proposed using a series of QPGs to separate spatially an incoming field into its TM components, each of which can be detected directly with a photodetector. (Huang and Kumar 2013) This obviates the need for coherent detection facilitated by the addition of a strong coherent local oscillator field, which creates added noise in the form of shot noise. This capability may open new possibilities for multistate quantum key distribution or for information transmission in photon-starved environments, such as deep-space communications. (Banaszek, *et al* 2019)

The measurement outcome of an ideal photodetector is proportional to the integrated *intensity* $\hat{E}_f^{(-)}\hat{E}_f^{(+)}$ of the field:

$$\hat{\mathcal{M}}(t) = \int_{-\infty}^{t} dt' \, R(t,t') \, \hat{E}_f^{(-)}(t') \hat{E}_f^{(+)}(t') \tag{31}$$

where $R(t,t')$ is the temporal response function of the detector to an impulsive input (we have suppressed the spatial dependence of the detector for clarity), and $\hat{E}_f^{(+)}(t')$ is the field filtered by a device such as a fast temporal shutter, a narrow spectral filter, or by a frequency converter or quantum memory that can select a particular spatial-temporal mode. This filtered field is related to the input field by:

$$\hat{E}_f^{(+)}(\tau) = \int ds \, \mathcal{G}(\tau,s) \, \hat{E}_{in}^{(+)}(s) \tag{32}$$

where $\mathcal{G}(\tau,s)$ is the response function of the linear filter device, related directly to the Green functions discussed earlier. Consider an input field of the form in Eq.(29):

$$\hat{E}_{in}^{(+)}(t) = \sum_j \hat{A}_j^\dagger u_j(t) \tag{33}$$

where the $u_j(t)$ are the natural modes of the signal as determined by the source physics.

How can a detector be made sensitive to the field in one isolated signal TM, say $u_J(t)$, while ignoring all other TMs? To achieve such mode isolation would require that the filter function be separable: $\mathcal{G}(\tau,s) \propto \psi_J(\tau) u_J(s)$, where at this point $\psi_J(\tau)$ could be an arbitrary function. This 'separability condition' is consistent with a filter function written as a Mercer expansion

$$\mathcal{G}(\tau,s) = \sum_j \lambda_j \psi_j(\tau) u_j(s) \, , \tag{34}$$

with only the $j = J$ coefficient being nonzero, that is $\lambda_j = \delta_{jJ}$. Typically, a linear filter is not separable in this way, however coherent filters such as pulsed frequency conversion or quantum memory can achieve it, leading to mode-selective detection, as discussed earlier. Such filters must be time-nonstationary, that is, time-varying relative to an external clock.

The most common filter types cannot be perfectly selective with respect to the TMs defined by a pulsed source of light. For example, consider a fast temporal shutter, which simply multiplies the signal by a gate function $g(\tau)$, constructed as:

$$\mathcal{G}(\tau,s) = g(\tau)\delta(\tau-s) \tag{35}$$

The filtered field is then

$$\hat{E}_f^{(+)}(\tau) = g(\tau)\sum_j \hat{A}_j^\dagger u_j(\tau) \tag{36}$$

In this case, the shutter will pass some portion of all signal modes, since all occupy the same time domain. The detector's mean output will then be (assuming a long integration time):

$$\langle \hat{\mathcal{M}} \rangle = \sum_{i,j} \langle \hat{A}_i^\dagger \hat{A}_j \rangle \int_{-\infty}^{\infty} d\tau\, R(t,\tau)|g(\tau)|^2\, u_i(\tau)u_j^*(\tau) \tag{37}$$

In the case that $R(t,\tau)|g(\tau)|^2$ is approximately constant over the duration of the TMs (so a shutter with a long opening window and a photodetector that responds much slower than the duration of the TMs themselves), the detector operator reduces to: $\hat{\mathcal{M}} = \sum_j \hat{A}_j^\dagger \hat{A}_j = \sum_j \hat{n}_j$. That is, the photodetector registers the total number of photons incident on it from all temporal modes.

A second example is a time-stationary spectral filter, such as a spectrometer, with $\mathcal{G}(\tau,s) = g(\tau-s)$. This situation is complementary to the shutter case, and the filter will pass some portion of all signal modes, since all occupy the same frequency domain.

One particularly important scenario is the characterization of single-photon pulses. The procedures used for this task lie at the heart of key protocols in, for example, quantum key distribution, and in linear optical operations that may be used in communications and simulation.

The way in which a single photon occupies the TM modes (or indeed any set of basis modes for the optical field) makes a difference to the outcome of measurements that are made on the light. Without careful design of the detector, certain properties (such as purity) may be imputed to the state of the light that are not in fact the case.

Consider, for instance, the common arrangement for measuring the properties of single photons: light is input to a 2-by-2 network (e.g., a beam splitter), as in Fig. 11(a), at the outputs of which are two photodetectors.

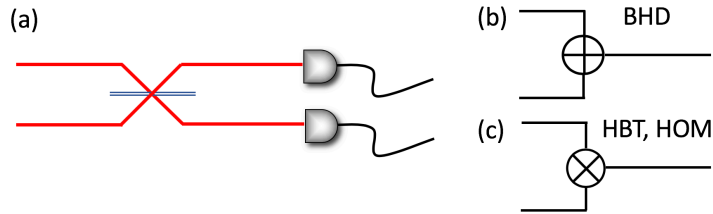

Fig. 11. (a) 2x2 network for characterization of quantum light. The photodetectors at the output resolve photon number in principle, although the resolution need not be at the single-photon level for some applications. (b) in Balanced Homodyne Detection (BHD) the registered signal is the difference in the photon numbers measured by the two photodetectors, when the quantum pulse enters one port of the network and a strong classical light pulse in a single TM enters the second. (c) in the Hanbury Brown and Twiss (HBT) and the Hong-Ou-Mandel (HOM) arrangements the signal is the product of the two photon numbers, proportional to the intensity correlation at the output of the network. For HBT one input port is empty, whereas for HOM both ports have quantum light inputs. Often these inputs are two independent single photons.

This arrangement is used in several different ways to determine the characteristics of quantum light beams. Indeed, the first experiments in which single TMs were measured made use of this device, in a configuration known as balanced homodyne detection (BHD). See Fig. 11(b). In that case, the quantum pulse is sent into one of the input ports, and a strong classical light pulse (the local oscillator) into the second. The measurement then consists of the difference in the photon numbers registered at each of the photodetectors, which can be represented by the operator

$$\hat{\mathcal{M}}_{BHD} \propto \sum_{j \in (1)} \hat{n}_j - \sum_{k \in (2)} \hat{n}_k \qquad (38)$$

where the first sum is over modes in port 1 and the second over those in port 2. The outcome of this measurement is a sample from the ensemble representing the in-phase amplitude of the light in the TM that matches exactly the temporal shape of the local oscillator pulse. Thus homodyne detection acts as a coherent filter for the TM of choice. (Raymer, *et al* 1995) The ability to isolate and detect light in a single TM enabled the first experiments demonstrating quantum state tomography. (Smithey, *et al* 1993) Tomography of quantum states is an important tool in quantum information science. (Lvovsky and Raymer 1990)

The same arrangement without any LO and the ancillary input port, and where the measurement is now the intensity correlation of the outputs, represented by the operator that is the product of the total photon number operators at each output port:

$$\hat{\mathcal{M}}_{co} \propto \sum_{j \in (1)} \hat{n}_j \sum_{k \in (2)} \hat{n}_k \qquad (39)$$

See Fig. 11(c). Note that this measurement operator assumes the detector can resolve photon numbers, though the experiment is often done with 'click' detectors, registering the presence or absence of photons. This arrangement is named for its inventors Hanbury-Brown and Twiss (HBT), who first used it for stellar intensity interferometry. The intensity correlation is now commonly used to test the purity of single photons. If a single-photon state to be tested is sent to one of the input ports of the network, with vacuum at the other, then the HBT measurement is proportional to $g_{HBT}^{(2)}(0)$ of the input light. This quantity is zero for single photons, since only one of the detectors will receive any light for such an input, so there can be no intensity at the other. The product of the two intensities is thus zero.

Since the measurement does not access coherences between the different photon-number sectors of the state impinging upon it, then without loss of generality, the state can be written for the purposes of this detection setup as a mixture:

$$\hat{\rho} = \sum_n p_n \hat{\rho}_n \qquad (40)$$

where $\hat{\rho}_n = \hat{a}_n^\dagger |vac\rangle\langle vac| \hat{a}_n$ is a single-photon state in TM *n* and $p_n$ is its probability. The measurement outcome in the single-photon case is $\langle \hat{\mathcal{M}}_{co} \rangle = Tr(\hat{\mathcal{M}}_{co} \hat{\rho}) = 0$, consistent with the expected HBT result, $g_{HOM}^{(2)} \propto \langle \hat{\mathcal{M}}_{co} \rangle = 0$.

However, in the case when two uncorrelated single photons are sent to two distinct beam-splitter inputs, the state is a direct product:

$$\hat{\rho} = \sum_{n \in (a)} p_n \hat{\rho}_n \otimes \sum_{m \in (b)} \pi_m \hat{\rho}_m \tag{41}$$

and the apparatus now measures the Hong-Ou-Mandel interference between the two input photons. The outcome of this measurement is (Mosley *et al* 2008b)

$$\langle \hat{\mathcal{M}}_{co} \rangle = Tr(\hat{\mathcal{M}}_{co} \hat{\rho}) = 1 - \sum_n p_n \pi_n \tag{42}$$

Thus, if the two input photons are in exactly the same TM, then the interference is complete and the outcome is consistent with $g^{(2)}_{HOM} = 0$. If however the photons are in different TMs, then the interference is incomplete, and the HOM "visibility" ($1 - g^{(2)}_{HOM}$) is reduced. Indeed, this is a signature of the so-called 'distinguishability' of the photons. But that is not the only important characteristic.

If, for instance, the photons are in identical states of more than one mode but are in mixed states, the interference visibility is reduced proportionally to the degree of mixedness (in fact the purity of each of the input photon states) so that $1 > g^{(2)}_{HOM} > 0$. The impurity of the state here is due entirely to the single photon occupying more than one mode, and not at all due to there being some possibility of more (or less) than one photon in each of the input states.

The important point to note is that the HBT and HOM measurements measure the purity of the state in two different degrees of freedom — 'mode space' versus 'number space'— and the latter is not simply a measure of the degree to which two photons are identical. Even two photons in exactly the same mixed state will not give perfect interference. The physical basis for this is simply that in a mixed state, the probability that the two photons are in exactly the same TM on any given run is less than one.

## 10. Summary, Conclusions

The concept of temporal modes (TMs) of the electromagnetic field, introduced by Roy Glauber and his colleagues, has proven to have significant utility both in fundamental quantum optics and in future quantum photonic applications. TMs are a useful and efficient way to describe non-stationary random processes, such as those arising in the quantum processes of light emission through both nonlinear optical scattering and fluorescence processes. The mode functions themselves depend on the process (in fact its field correlation function) and so are optimally adapted to the underlying physics, and the field is constructed from these functions with weights that are uncorrelated random variables.

Because TMs form an orthogonal and complete set, they are also useful for coding information. In fact, since they overlap both in time and frequency, they are the most compact way of packing information into a particular domain of chronocyclic (time-frequency) phase space. A primary application may be in cases were the signals being transmitted is in the 'photon-starved' regime, such as future embodiments of deep-space communication as from a distant spacecraft to Earth. (Banaszek, *et al* 2019)

Making full use of TMs poses some questions and challenges. These include discovering: which applications truly benefit from using TMs; how to multiplex and demultiplex large numbers of TMs; how to create arbitrary superpositions of them across multiple Fock subspaces of the quantum field state space; and how to carry out such operations efficiently in an integrated-optics platform and across a wide range of wavelengths.

In this article, we have sketched a brief history of how the TM concept has been applied to pulsed quantum processes, following Glauber's original idea. Further, we have shown how the ideas have a broad application to many varieties of quantum light sources, including superfluorescence, stimulated Raman scattering, parametric downconversion and four-wave mixing. This broad range indicates the wide applicability of the concept, as well as its practical uses.

Making the most of TMs demands a means to generate them, manipulate them and detect them. The ability to change one into an arbitrary superposition of others and back again is critical. We have described the physics of quantum light sources, some of which can be used to generate states of a specified superposition of TMs. These sources can be adapted to give 'pure' single photons in a single designed TM. Recent research has shown how the TMs can be demultiplexed and manipulated by a device called a 'quantum pulse gate,' which can be implemented in various different ways using nonlinear optical processes. Finally, TMs can be detected, and indeed fully characterized, by means of suitable arrangements of photodetectors (which are not intrinsically TM sensitive) together with QPGs, filters and ancillary light beams, including classical ones. We have given a model for detecting TMs using photon-number-resolving detectors, showing how mixedness in both modes and in photon number impacts a number of common measurement configurations.

The burgeoning field harnessing TMs for useful applications draws on the long history of these fascinating objects, and in that sense, Glauber's legacy continues. As we move toward broadband quantum networking using light, we anticipate that TMs will become of increasing utility.

**Acknowledgements**

The authors' work was in many cases inspired by the contributions of Roy Glauber and his collaborators. We thank Fritz Haake for permission to use Figure 1 (b). MR was supported in part by the U.S. National Science Foundation (grants PHY-1820789 and PHY-1839216); and IW by the Networked Quantum Information Technologies Hub (NQIT) as part of the UK National Quantum Technologies Programme (Grant EP/N509711/1), as well as the ERC Advanced Grant MOQUACINO.

29(Zeldovich and Klyshko 1996) Y. Zeldovich and D. Klyshko, 1969. *JETP LETTERS-USSR, 9*, 40 (1969).

(Zhu, *et al* 2007) Zhu, Z., Gauthier, D.J. and Boyd, R.W., 2007. Stored light in an optical fiber via stimulated Brillouin scattering. *Science*, *318*(5857), pp.1748-1750.
29